\documentclass[useAMS,usenatbib]{mn2e}
\usepackage[pdftex]{graphicx}
\usepackage[pdftex,usenames,dvipsnames]{xcolor}
\usepackage{amssymb}
\usepackage{natbib}
\usepackage{aas_macros}
\usepackage{url}
\usepackage[none]{hyphenat}
\usepackage{times}
\usepackage[linktocpage=true,backref=true,pagebackref=false,bookmarks=true,bookmarksnumbered=False,colorlinks=true,linkcolor=NavyBlue,citecolor=NavyBlue,urlcolor=NavyBlue]{hyperref}
\usepackage[all]{hypcap}
\pdfpageattr {/Group << /S /Transparency /I true /CS /DeviceRGB>>}  

\title[Circinus dust lane]{The nuclear dust lane of Circinus: collimation without a torus}

\author[Mezcua et al.]
{M. Mezcua$^{1,2,3}$\thanks{Email: marmezcua.astro@gmail.com}, M. A. Prieto$^{3,4}$, J.A. Fern\'andez-Ontiveros$^{5}$, K. R. W. Tristram$^{6}$\\ 
$^1$D\'epartement de Physique, Universit\'e de Montr\'eal, C.P. 6128, Succ. Centre-Ville, Montreal, Quebec H3C 3J7, Canada\\
$^2$Harvard-Smithsonian Center for Astrophysics (CfA), 60 Garden Street, Cambridge, Massachusetts 02138, USA\\
$^3$Instituto de Astrof\'isica de Canarias (IAC), E--38200 La Laguna, Tenerife, Spain \\
$^4$Universidad de La Laguna, Dept. Astrof\'isica, E--38206 La Laguna, Tenerife, Spain\\
$^5$Istituto di Astrofisica e Planetologia Spaziali (INAF-IAPS), Via Fosso del Cavaliere 100, 00133 Roma, Italy\\  
$^6$European Southern Observatory, Alonso de C\'ordova 3107, Vitacura, Casilla 19001, Santiago de Chile, Chile\\
}

\date{}

\pagerange{\pageref{firstpage}--\pageref{lastpage}} \pubyear{2013}

\def\LaTeX{L\kern-.36em\raise.3ex\hbox{a}\kern-.15em
    T\kern-.1667em\lower.7ex\hbox{E}\kern-.125emX}

\begin{document}

\label{firstpage}

\maketitle

\begin{abstract}
In some AGN, nuclear dust lanes connected to kpc-scale dust structures provide all the extinction required to obscure the nucleus, challenging the role of the dusty torus proposed by the Unified Model. In this letter we show the pc-scale dust and ionized gas maps of Circinus constructed using sub-arcsec-accuracy registration of infrared VLT AO images with optical \textit{Hubble Space Telescope} images. We find that the collimation of the ionized gas does not require a torus but is caused by the distribution of dust lanes of the host galaxy on $\sim$10 pc scales. This finding questions the presumed torus morphology and its role at parsec scales, as one of its main attributes is to collimate the nuclear radiation, and is in line with interferometric observations which show that most of the pc-scale dust is in the polar direction. We estimate that the nuclear dust lane in Circinus provides $1/3$ of the extinction required to obscure the nucleus. This constitutes a conservative lower limit to the obscuration at the central parsecs, where the dust filaments might get optically thicker if they are the channels that transport material from $\sim$100 pc scales to the centre.

\end{abstract}


\section{Introduction}
 Supermassive black holes are ubiquitous at the centre of galaxies (\citealt{2013ApJ...764..184M}), but only a fraction of them are actively accreting in the form of an active galactic nucleus (AGN). Those supermassive black holes powering an AGN are thought to be surrounded on scales of a few parsec by a toroidal distribution of warm molecular gas and dust (hence called ``dusty torus'') that constitutes the reservoir of material that fuels the central engine. According to the Unified Model of AGN (\citealt{1988ApJ...329..702K}; \citealt{1993ARA&A..31..473A}; \citealt{1995PASP..107..803U}) the dusty torus is also responsible for the AGN dichotomy into type 1 (face-on view, unobscured nuclei with broad and narrow optical emission lines) and type 2 (edge-on view, obscured nuclei with only narrow optical emission lines) and for collimating the ionizing radiation. Evidence for this scenario relies on two compelling results: (i) spectro-polarimetric observations which reveal the presence of broad emission lines in some type 2 AGN (e.g. \citealt{1985ApJ...297..621A}; \citealt{1994ApJ...430..196K}) and (ii) the observed anisotropy of the nuclear ionizing radiation field giving rise to conical-shape morphologies of the narrow-line regions with the cone apex at the nucleus (e.g. \citealt{1989Natur.341..422T}; \citealt{1993ApJ...418..668P}). In the obscuration/reflection scenario of \cite{1985ApJ...297..621A}, the ionization cones result from shadowing of the nucleus by an optically-thick pc-scale torus. Many type 2 nuclei are Compton thick in the X-rays, a result often interpreted as further evidence for a pc-scale torus (e.g. \citealt{1999ApJ...522..157R}). However, not always do type 2 AGN present large X-ray column densities ($N_\mathrm{H}>10^{22}$ cm$^{-2}$; e.g. \citealt{2009MNRAS.398.1951P}; \citealt{2012MNRAS.426.3225B}; but see \citealt{2015arXiv151105566B}) and high $N_\mathrm{H}$ can be caused by the host galaxy seen edge-on (e.g. \citealt{2014MNRAS.438..647H}).
 
The torus is expected to be  of parsec scales  with a clumpy morphology and be made of cold molecular material, a large fraction of it being dust particles that due to their proximity to the nucleus  will absorb the UV emission from the accretion disc and re-radiate it in the infrared (IR; \citealt{1998agn..book.....K}). Current IR interferometric observations now feasible for the nearest AGN have the capability to resolve this structure (e.g. \citealt{2013A&A...558A.149B}). Instead of the expected geometrically-thick torus, interferometric observations reveal the presence of a mid-IR pc-scale polar elongation and a disc-like component perpendicular to the system axis (i.e. in NGC\,1068, \citealt{2004Natur.429...47J}, \citealt{2014A&A...565A..71L}); NGC\,424, \citealt{2012ApJ...755..149H}; NGC\,3783, \citealt{2013ApJ...771...87H} and Circinus, \citealt{2014A&A...563A..82T}). Extended mid-IR emission in the polar direction has been also observed on scales $>$ 100 pc in the Circinus galaxy and a few other objects \mbox{(e.g. \citealt{2005ApJ...618L..17P}; \citealt{2010A&A...515A..23H}; \citealt{2010MNRAS.402..879R}). }
Although a radiatively-driven dusty wind could explain such polar-elongated dust structures (\citealt{2012ApJ...755..149H,2013ApJ...771...87H}), these findings pose a further challenge to the role of the torus in the framework of the Unified Model and how this disc-like component plus extended polar dust is able to account for the obscuration -viewing angle dependent- and the morphology of the ionized gas.

In this letter we provide direct observational evidence that a dusty torus is not required to explain the collimation of the conical-like 
ionized gas structure seen in Circinus. Circinus is an almost edge-on ($\sim$65 deg) SA(s)b galaxy, considered to be a prototype Seyfert 2 AGN with a narrow emission line spectrum (\citealt{1994A&A...288..457O}), polarized broad emission lines (\citealt{1998A&A...329L..21O}) and a one-sided ionization cone that extends up to kpc scales in H$\alpha$ and [OIII] (e.g. \citealt{1997ApJ...479L.105V}), $\sim$60 pc in X-rays (\citealt{2001ApJ...557..180S}) and $\sim$30 pc in the coronal line [SiVII] 2.48 $\mu$m (\citealt{2005MNRAS.364L..28P}). The nucleus is Compton-thick ($N_\mathrm{H}\sim4\times10^{24}$ cm$^{-2}$; \citealt{1999A&A...341L..39M}) and only visible at wavelengths $\lambda >1.6 \mu$m (\citealt{2004ApJ...614..135P}). At a distance of $\sim$4 Mpc (\citealt{1977A&A....55..445F}), 1 arcsec corresponds to $\sim$19 pc.

\section{Data and Analysis}
\subsection{IR adaptive optics and \textit{HST} optical data}
Circinus was observed with the adaptive optics (AO) assisted near-IR instrument Naos-Conica (NaCo; \citealt{2003SPIE.4839..140R}) at the ESO Very Large Telescope (VLT) as part of a larger sample of near and bright AGN. The sample and data reduction have been described elsewhere (\citealt{2010MNRAS.402..724P}; \citealt{2010MNRAS.402..879R}; \citealt{2014MNRAS.442.2145P}; \citealt{2015MNRAS.452.4128M}). For the purpose of this Letter, we use the $2\, \rm{\micron}$ broad-band \textit{Ks} filter ($\lambda_c = 2.180\, \rm{\micron}$, $\Delta\lambda = 0.350\, \rm{\micron}$) to obtain the near-IR continuum image, as at this band a maximum contrast between the AGN and the host galaxy is observed (e.g. \citealt{2010MNRAS.402..724P,2014MNRAS.442.2145P}). The NaCo beam resolution in the Ks-band is of 0.16 $\pm$ 0.02 arcsec.

\begin{figure*}
 \includegraphics[width=\textwidth]{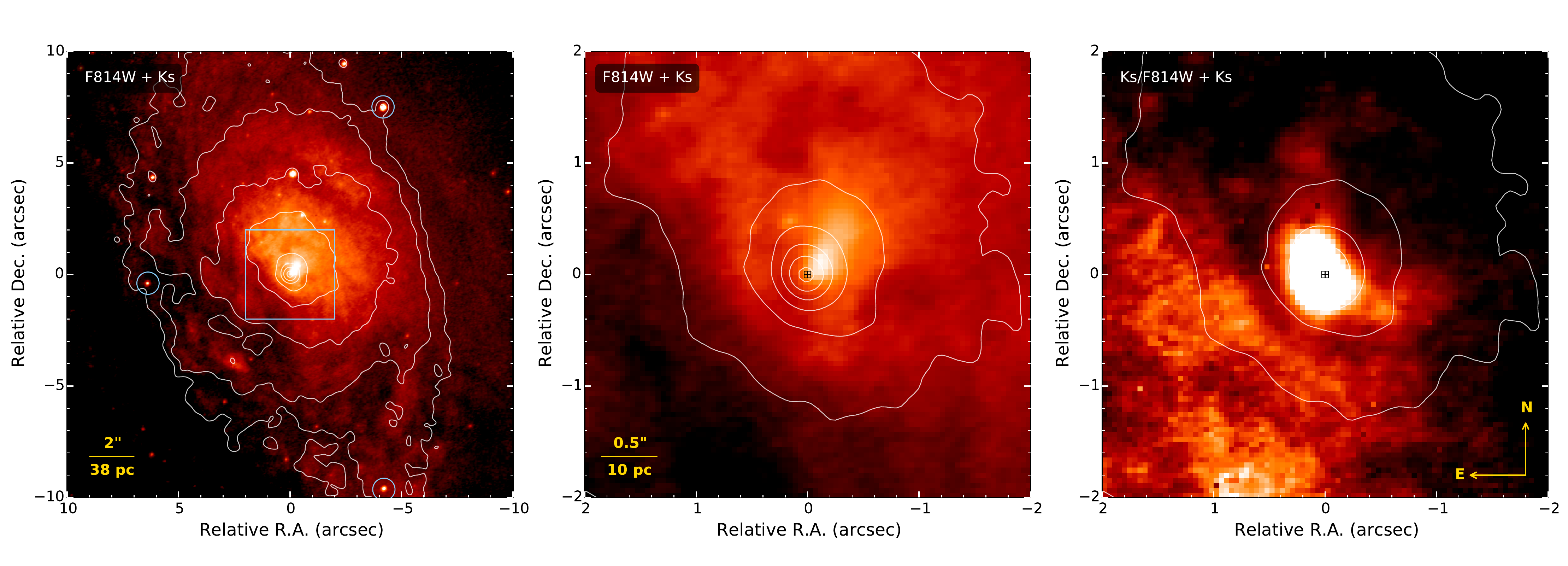}
 \caption{\textbf{Left}: \textit{HST/F814W} image with NaCo/\textit{Ks}-band continuum contours in white. The FoV is 20 arcsec $\times$ 20 arcsec. Blue circles mark the position of the point-like sources used for image alignment. The inner  4 arcsec $\times$ 4 arcsec region (blue square) is shown in detail in the next panels. \textbf{Middle}: \textit{HST/F814W} image with NaCo/\textit{Ks}-band continuum contours in white. \textbf{Right}: \textit{Ks}/F814W flux ratio with NaCo/\textit{Ks}-band continuum contours in white. The brighter regions in the \textit{Ks}/F814W image denote higher extinction and hence dust absorption. The position of the nucleus and its error is marked with a cross in the middle and right panels. The angular resolution in all panels is that of the NaCo \textit{Ks}-band (0.16 arcsec).}
 \label{fig1}
\end{figure*}

We use the archival \textit{Hubble Space Telescope} (\textit{HST}) broad \textit{I}-band image from the \textit{F814W} filter of the WFPC2\footnote{Wide-Field Planetary Camera 2.} to construct an extinction map \textit{Ks}/\textit{I} or dust map. When there is dust at a location, \textit{I} is suppressed there whereas \textit{Ks} is enhanced relative to other locations where there is less or no dust. These enhanced regions show brighter in our \textit{Ks}/\textit{I} figure, indicating higher extinction $A_{V}$, and take the shape of filaments and lanes, suggesting these are dust regions. The \textit{HST/F814W} image corresponds to the reddest \textit{HST} optical image available and provides a large number of common reference sources with the NaCo/Ks-band image to be used for the image alignment. The registration of the NaCo/Ks and \textit{HST/F814W} images is reported in \cite{2015MNRAS.452.4128M} and was performed using the centroid position of three reference sources (the nucleus was not used; see Fig.~\ref{fig1}, top left panel) to derive the relative offsets between the images. The alignment uncertainty is of 30 mas. 
Using the same procedure, we register the WFPC2/\textit{F502N}, WFPC2/\textit{F656N} and NICMOS/\textit{F187N} filter images and their corresponding adjacent continuum to the NaCo/Ks and \textit{HST/F814W} images. Five common sources are used in this image registration. The alignment uncertainties are of 10 mas, 40 mas and 20 mas for the WFPC2/\textit{F502N}, WFPC2/\textit{F656N} and NICMOS/\textit{F187N} alignments, respectively. The [\textsc{O\,iii}], H$\alpha$+[\textsc{N\,ii}] and Pa$\alpha$ images are then constructed by subtracting the adjacent continuum following standard pure-line extraction procedures. 


\subsection{Radio observations: the pc-scale radio jet}
We searched for sub-arcsec resolution archival radio data of Circinus with the aim of studying the morphology and orientation of its radio jet at pc scales. We re-reduced archival 22 GHz data obtained in March 2005 with the Australia Telescope Compact Array (ATCA) using the MIRIAD package following standard procedures. The data were then imaged with robust weighting, resulting in a 0.37 arcsec $\times$ 0.24 arcsec beam oriented at a position angle (P.A.) of -8 deg. The core of the radio jet of Circinus (central component in Fig.~\ref{fig2}, bottom right) has a flux of 364.5 $\pm$ 4.9 mJy. The source is slightly resolved, with a size of 0.74 arcsec $\times$ 0.52 arcsec (measured from the 5$\sigma$ contour) with an extension toward the North-West along a P.A. of about -80 deg. This P.A. is consistent with the direction of the polar emission detected by mid-IR interferometry (P.A. = -76 deg North to East) and roughly perpendicular to the nuclear dust emitter (P.A. = 46 deg; \citealt{2014A&A...563A..82T}). Six more blobs are detected above a 5$\sigma$ level and are most likely part of the large-scale radio lobes detected at lower frequencies (e.g. \citealt{1998MNRAS.297.1202E}).

\section{Results and Discussion}
\subsection{The location of the nucleus}
\label{nucleus}
The position of the nucleus can be identified as an outstanding point-like source in the NaCo $2\, \rm{\micron}$ \textit{Ks}-band image, where it reaches its maximum contrast. This is particularly important for those galaxies hosting a type 2 AGN, in which the nucleus is obscured either by a dust filament crossing the nuclear region (\citealt{2010MNRAS.402..724P,2014MNRAS.442.2145P}) or a pc-scale torus. As a consequence, an offset between the \textit{Ks}-band and optical peaks of emission of several tens of pc is observed. Such IR--optical shift was found for some Sy2 galaxies (e.g. NGC\,1068, NGC\,1386 and NGC\,7582; \citealt{2014MNRAS.442.2145P}), including Circinus (\mbox{\citealt{2004ApJ...614..135P}}; \citealt{2015MNRAS.452.4128M}).  
Here we unambiguously identify the position of the nucleus of Circinus, revealed as a bright point-like source in the $2\, \rm{\micron}$ \textit{K}-band images, with a precision of $30\, \rm{mas}$. This near-IR nucleus is found to be shifted by $160\, \rm{mas} \pm 30\, \rm{mas}$ to the peak of \textit{I}-band emission, in agreement with the $0\farcs15$ offset derived by \cite{2004ApJ...614..135P} and indicating that the nucleus is clearly obscured at optical wavelengths (see Fig.~\ref{fig1}, middle and right panels).

\begin{figure*}
 \includegraphics[width=\textwidth]{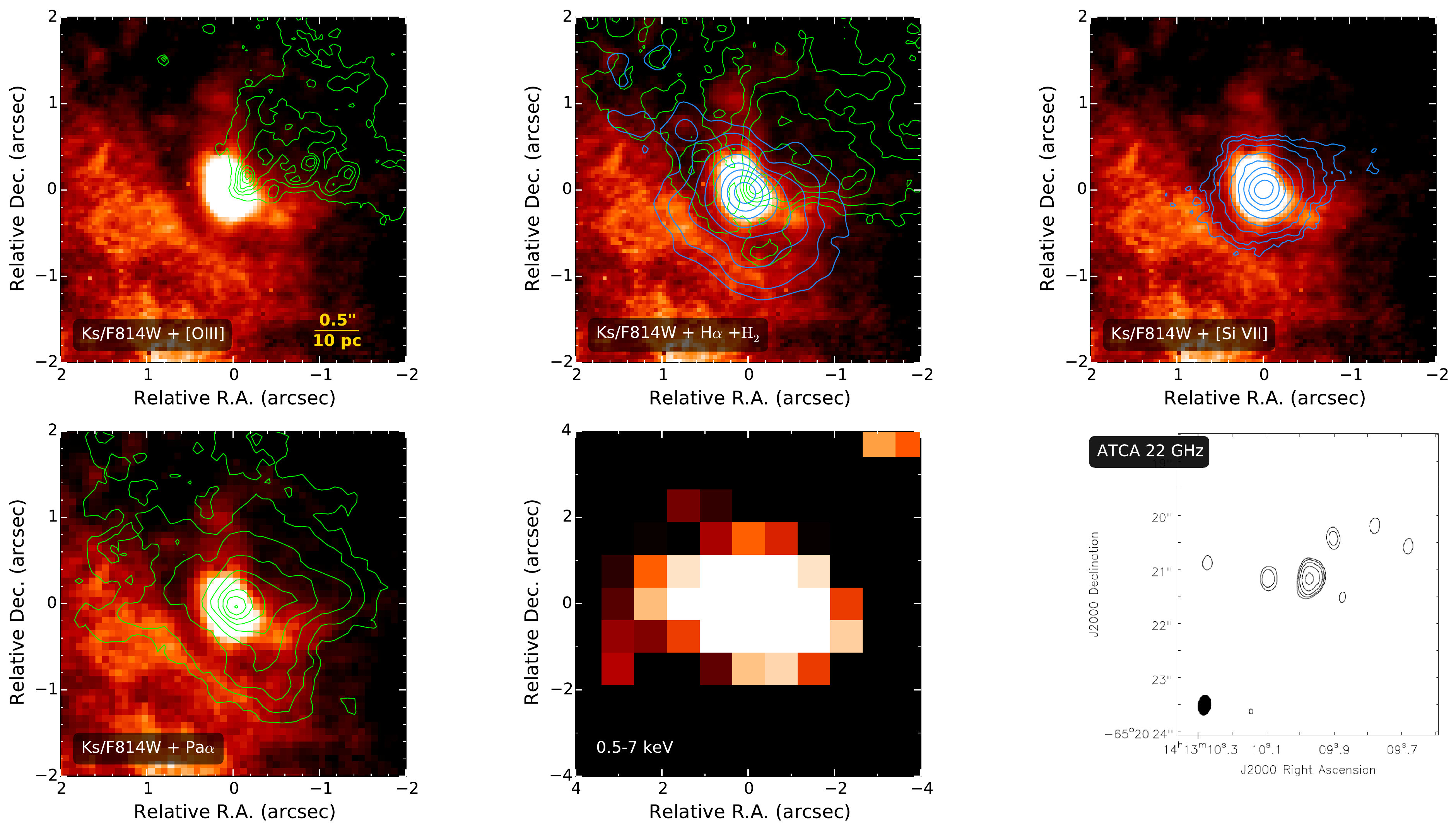}
 \caption{\textit{Ks}/F814W flux ratio with emission line contours of [\textsc{O\,iii}] (\textbf{top left}), H$\alpha$ in green and warm $H_{2}$ molecular gas in blue (\textbf{top middle}), \textsc[{Si\,vii}] (\textbf{top right}), and Pa$\alpha$ (\textbf{bottom left}). The lowest [\textsc{O\,iii}], H$\alpha$, $H_{2}$, and Pa$\alpha$ contours correspond to $5\sigma$, while for \textsc[{Si\,vii}] the lowest contour corresponds to 2$\sigma$ (to better match the map of \citealt{2004ApJ...614..135P}). The FoV is 4 arcsec $\times$ 4 arcsec. The brighter regions in the \textit{Ks}/F814W images denote higher extinction and hence dust absorption. The angular resolution of the \textit{Ks}/F814W images is 0.16 arcsec, that of $H_{2}$ is 0.16 arcsec (\citealt{2015MNRAS.452.4128M}) and that of \textsc[{Si\,vii}] is 0.19 arcsec (\citealt{2005MNRAS.364L..28P}).
 \textbf{Bottom middle}: \textit{Chandra} X-ray image in the 0.5-7 keV band with a FoV of 8 arcsec $\times$ 8 arcsec. \textbf{Bottom right}. ATCA radio continuum map at 22 GHz. The synthesized beam has a FWHM of 0.37 arcsec $\times$ 0.24 arcsec at a P.A. of -8 deg. Contours are plotted as (5, 10, 20, 40, 80) times the off-source rms of 3.3 mJy beam$^{-1}$. North is up and East is to the left.}
 \label{fig2}
\end{figure*}

\subsection{Collimation by dust lanes}
\label{collimation}
The nucleus (or $2\, \rm{\micron}$ peak of emission) is located behind a dust filament or lane that extends up to $\sim$1.5 arcsec ($\sim$30 pc; see Fig.~\ref{fig1} right panel) forming a horned shape. This horn-shaped filament appears to delimit the sharp edges of the [\textsc{O\,iii}] cone of emission as well as the main structure of the cone of H$\alpha$ emission (Fig.~\ref{fig2}). Additional clumpy H$\alpha$ emission is detected at a 5$\sigma$ level outside the cone defined by the horn-shaped filament. This clumpy H$\alpha$ emission most likely traces low-level star formation relatively unobscured by foreground dust, as supported by its spatial coincidence with clumpy warm molecular 
H$_{2}$ 1-0 S(1) 2.12 $\mu$m line gas emission Northeast of the nucleus (Fig.~\ref{fig2}, top middle; \citealt{2015MNRAS.452.4128M}). The sharp definition of the edges of the [\textsc{O\,iii}] cone (Fig.~\ref{fig2}, top left) being strictly traced at the vortex region by the horn-shaped dust lanes indicates that the ionizing radiation is not necessarily collimated by a nuclear torus but that dust at scales of a few tens of pc is responsible for the observed cone-like morphology. If this is the case, a more extended distribution should be observed in H$\alpha$ which is less affected by dust extinction, as it is indeed the case. An even more isotropic distribution is expected and indeed observed when the ionized gas is traced at the same $\sim$30 pc scale by much redder emission lines that are not so much affected by dust extinction (e.g. Pa$\alpha$ or \textsc[{Si\,vii}] 2.48 $\mu$m). For example, the emission in \textsc[{Si\,vii}] is rather isotropic extending in rather symmetric form in all directions (Fig.~\ref{fig2}, top right). The Pa$\alpha$ emission (Fig.~\ref{fig2}, bottom left) is even more isotropic than H$\alpha$ and [\textsc{O\,iii}], reinforcing that the latter are strongly shaped by dust extinction at a few tens of pc. 
Both the nuclear X-ray emission (Fig.~\ref{fig2}, bottom middle; see also \citealt{2001ApJ...557..180S}) and the pc-scale radio jet (Fig.~\ref{fig2}, bottom right) also extend along the directions of the ionized gas. The X-ray emission may tentatively be ascribed to the horn-shaped dust filaments. Still, the angular resolution in these two spectral regions is insufficient to set up a preferential direction or morphology.

\subsection{Extinction}
To test to which degree the nuclear dust lanes contribute to the obscuration of the nucleus, we estimate the extinction $A_V$ produced by these dust lanes by measuring the galaxy colours in several apertures (of 0.1 arcsec radius) in the dusty structure and comparing them to dust-free regions (of aperture radius 0.1--0.2 arcsec) located at a distance of $\sim$3 arcsec ($\sim$ 60 pc) from the centre. We note that these dust-free regions may still be affected by dust extinction as we are measuring at the lowest detection level of the detector, thus we are deriving a lower limit on the $A_V$. The extinction law from \cite{1989ApJ...345..245C} is applied. 
On the assumption of a foreground dust screen, which is the most appropriate dust distribution for the filaments, an extinction of $A_V\gtrsim$ 3 mag is obtained in the dusty structure at a distance $\sim$1 arcsec ($\sim$19 pc) from the nucleus and $A_V\gtrsim$ 6.9 mag closer to the nucleus (at a distance of 0.2 arcsec, $\sim$4 pc). This value is in agreement with previous extinction estimates, also assuming a foreground dust screen: close to the nucleus, a lower limit of $A_V$ = 6 is derived from \textit{J-K} extinction maps (\citealt{2004ApJ...614..135P}); $A_V$ = 6.3 mag is measured within the central 1.5 arcsec of the stellar component by \cite{1998ApJ...493..650M}.
It should be noted that with this method we cannot measure the extinction directly on the nucleus as the colours at the nucleus locations and immediate surrounding relate to dust emitting at about 2 $\mu$m heated by the AGN. The location of our $A_V$ measurements is a compromise between being the nearest to the centre and still far for the dust to be heated at the right temperatures to emit at 2 $\mu$m; therefore, the \textit{I-K} colours should reflect just dust extinction.

Taking $A_V\gtrsim$ 6.9 mag as a face value, we check next whether this amount of extinction is enough to hide the nucleus of Circinus in the optical. To that aim, we scale the Seyfert 1 template from \cite{2010MNRAS.402..724P}, also based on subarcsec-resolution observations, to the mid-IR flux of Circinus measured in the same work ($17.6\, \rm{Jy}$ at $18.72\, \rm{\micron}$). Assuming a foreground extinction layer, we estimate that an $A_V \sim 20\, \rm{mag}$ would be needed in order to obscure the Seyfert 1 template below the upper limit of $1.6\, \rm{mJy}$ in the \emph{J}-band derived by \cite{2010MNRAS.402..724P}, in agreement with the silicate optical depth from \cite{2007A&A...474..837T} and in line with the lower limit of $A_V$ = 27.2 found by \cite{2015arXiv151105566B}. Thus, the dust traced in our maps is able to explain $\gtrsim 1/3$ of the extinction needed to turn a bright Seyfert 1 nucleus into a Seyfert 2 like Circinus. A thicker absorber at the centre is therefore needed to completely obscure the nucleus and produce the observed spectral energy distribution (SED). Still, as indicated before, our $A_V$ estimate is a conservative lower limit; the dust filaments may get thicker as we approach the centre, in particular if they trace material flowing towards the centre. We see compelling evidence for this flow of material through dust filaments and lanes in other nearby AGN (\citealt{2014MNRAS.442.2145P}; Nadolny et al. in preparation). Should that be the case, the dust filaments in Circinus must cast higher $A_V$ at the centre. The interferometric mapping of the central pc indicates already
a strong gradient in the silicate-derived extinction within the 2 pc (the average extinction agrees with our estimate from the SED): an increasing extinction is seen from the NW to the SE close to a P.A. = -90 deg (\citealt{2014A&A...563A..82T}), i.e. in line with the Western horn-shaped filament.

\subsection{Where is the dusty torus?}
We have seen that in the Circinus galaxy the collimation of the ionizing radiation does not require a pc-scale dusty torus. Instead, a larger dust distribution that extends up to $\sim$30 pc is responsible for collimating the ionizing radiation and hence for shaping the cone-like distribution of the ionized gas, which is by definition one of the main attributes of the torus. This larger dust distribution manifests itself in the form of a nuclear horn-shaped structure in our extinction maps. This result, together with the result from mid-IR interferometry that most of the pc-scale mid-IR dust emission extends in the polar rather than in the perpendicular direction to the cone (\citealt{2014A&A...563A..82T}), questions the presumed torus (donut-like) morphology, at least at pc scales. 
A much higher obscuration than that inferred from the horn-shape dust filaments is still needed to explain the high levels of obscuration and the detection of broad emission lines in polarized light. The nucleus of Circinus at 2 and 5 $\mu$m resolves into a slightly elongated source with a size of 1.9 $\pm$ 0.6 pc  (\citealt{2004ApJ...614..135P}). This size is in line with that of the interferometric polar direction component measured at 10 $\mu$m (\citealt{2014A&A...563A..82T}) and is also in broad agreement with our measured shift of 3 $\pm$ 0.6 pc between the optical and the IR peaks (Section~\ref{nucleus}). This shift could be interpreted as the size of the nuclear obscurer. The direction of the shift is coincident with that defined by the interferometric polar dust component, but the outer region of the absorber may have escaped detection by the interferometric observations if it gets progressively less optically thick outwards. \cite{2014A&A...563A..82T} interpret the polar dust component as the true cause of Circinus nuclear obscuration and call it the torus. We broadly agree with the interpretation, but associate the obscurer with the innermost parts of the large-scale ($\sim$30 pc) dust filaments crossing the nucleus, the horn shape, for the following reasons: 1) the filaments shape the morphology of the ionization cone (Section~\ref{collimation}); 2) their width, $\sim$0.2 arcsec (4 pc; Fig.~ \ref{fig2}), just fit in the range of the absorber size discussed above; 3) they could naturally account for the nuclear obscuration if getting thicker toward the inner parsecs, which is presumably the case given the strong extinction gradient measured by the interferometric observations. In addition, no temperature gradient is observed down to 0.1 pc (\citealt{2014A&A...563A..82T}), which can be easily accommodated with a large-scale obscurer, filament-like, instead of a torus. Still, the spatially resolved nuclear structure seen at 2 $\mu$m, of 2 pc, indicates that some fraction of the obscurer gets heated up to high temperatures as approaching the centre. 

The results showed in this Letter therefore explain how an ionization cone of a few parsecs in size can be shaped by the dust distribution of several tens of parsec of the host galaxy, instead of a pc-scale geometrically-thick toroidal structure, demonstrating that the nuclear obscuration responsible for the type 1/2 AGN dichotomy does not necessarily arise from the same structure that collimates the ionizing radiation. If confirmed for a significantly larger sample -as suggested by additional cases like NGC\,3169 or NGC\,7582 (\citealt{2014MNRAS.442.2145P})- this might imply a revision of the dust distribution in the torus.

\section*{Acknowledgements}
The authors thank the referee, L. Burtscher, for his valuable comments which have helped improve this manuscript, and M. Elvis for insightful discussion. MM acknowledges support from the NASA \textit{Chandra} Grant G05-16099X.

\bibliographystyle{mn2e} 
\bibliography{/Users/mmezcua/Documents/referencesALL}

\label{lastpage}

\end{document}